\documentclass{article}
\usepackage{spconf}

\usepackage{amsmath, amsfonts, amssymb, float, enumerate, bm, graphicx, mathtools, mathrsfs}

\usepackage{tikz}
\usetikzlibrary{arrows}
\usepackage{verbatim}
\tikzstyle{int}=[draw, fill=white!20, minimum size=2em]
\tikzstyle{init} = [pin edge={to-,thin,black}]

\restylefloat{table}
\newcommand{\Ihat}{\widehat{I}}

\newcommand{\curly}[1]{\left\{#1\right\}}
\newcommand{\round}[1]{\left(#1\right)}

\newcommand{\util}{\mathscr{U}}
\newcommand{\best}{\mathscr{B}}

\newcounter{eg}[section]
\renewcommand{\theeg}{\arabic{section}.\arabic{eg}}
\newenvironment{examp}[1][]{\refstepcounter{eg}
\par\medskip \noindent
   \textit{Example~\theeg. #1} \rmfamily}{\hfill $\square$   \hspace{-4.5pt} \vspace{6pt}}

\usepackage{amsmath, amssymb, bbm, xspace}
\usepackage{epsfig}
\usepackage{longtable}
\usepackage{color}
\usepackage{mathrsfs}
\usepackage{comment}
\usepackage{ifthen}
\newboolean{showcomments}
\setboolean{showcomments}{true}
\usepackage{courier}

\newtheorem{theorem}{Theorem}[section]

\newtheorem{corollary}[theorem]{Corollary}

\newtheorem{definition}{Definition}[section]

\def\bkE{{\rm I\kern-.17em E}}
\def\bk1{{\rm 1\kern-.17em l}}
\def\bkD{{\rm I\kern-.17em D}}
\def\bkR{{\rm I\kern-.17em R}}
\def\bkP{{\rm I\kern-.17em P}}

\def\bkZ{{\bf{Z}}}

\def\bkE{{\rm I\kern-.17em E}}
\def\bk1{{\rm 1\kern-.17em l}}
\def\bkD{{\rm I\kern-.17em D}}
\def\bkR{{\rm I\kern-.17em R}}
\def\bkP{{\rm I\kern-.17em P}}

\makeatletter
\newcommand{\pushright}[1]{\ifmeasuring@#1\else\omit\hfill$\displaystyle#1$\fi\ignorespaces}
\newcommand{\pushleft}[1]{\ifmeasuring@#1\else\omit$\displaystyle#1$\hfill\fi\ignorespaces}
\makeatother

\def\bkZ{{\bf{Z}}}
\def\b12{(\beta_1,\beta_2)}

\newenvironment{example}{{\noindent \bf Example}}{\hfill $\square$\hspace{-4.5pt}\vspace{6pt}}
\newcounter{example}
\renewcommand{\theexample}{\thesection.\arabic{example}}

\newcounter{remark}
\renewcommand{\theremark}{\thesection.\arabic{remark}}

\newenvironment{remark}{{\noindent \it Remark: }}{\hfill $\square$}

\def\Xscr{\mathcal{X}}

\newlength{\noteWidth}
\long\def\notes#1{\ifinner
{\tiny #1}
\else
\marginpar{\parbox[t]{\noteWidth}{\raggedright\tiny #1}}
\fi\typeout{#1}}

 \def\notes#1{\typeout{read notes: #1}} 

\newcommand{\wi}[1]{\widehat{#1}}

\DeclareMathOperator{\image}{Im}

\newcommand{\ie}{i.e.\@\xspace} 

\newcommand{\Real}{\ensuremath{\mathbb{R}}}

\def\rarr{\rightarrow}

\def\Pbb{{\mathbb{P}}}
\def\Nbb{{\mathbb{N}}}

\def\spose#1{\hbox to 0pt{#1\hss}}

\def\text #1{\hbox{\quad#1\quad}}

\def\xhat{{\hat x}}

\def\nthinsp{\mskip -2   mu}

\def\superstar{^{\raise 0.5pt\hbox{$\nthinsp *$}}}
\def\SUPERSTAR{^{\raise 0.5pt\hbox{$*$}}}

\def\lamstarT {\lambda^{\raise 0.5pt\hbox{$\nthinsp *$}T}}

\def\Fscr{{\cal F}}
\def\Dscr{{\cal D}}

\def\Pscr{{\cal P}}

\def\Xscr{{\cal X}}

\def\gtilde{\skew{4.5}\widetilde g}

\def\Ibar{\skew5\bar I}
\def\Itilde{\widetilde I}

\def\stilde{\widetilde s}

\def\xbar{\skew{2.8}\bar x}
\def\xhat{\skew{2.8}\widehat x}

\def\yhat{\skew3\widehat y}

\def\non{\nonumber}

\let\forallnew\forall
\renewcommand{\forall}{\forallnew\ }
\let\forall\forallnew

		\def\bkE{{\rm I\kern-.17em E}}
		\def\bk1{{\rm 1\kern-.17em l}}
		\def\bkD{{\rm I\kern-.17em D}}
		\def\bkR{{\rm I\kern-.17em R}}
		\def\bkP{{\rm I\kern-.17em P}}
		\def\bkY{{\bf \kern-.17em Y}}
		\def\bkZ{{\bf \kern-.17em Z}}
		\def\bkC{{\bf  \kern-.17em C}}

{\begin{list}{}%
         {\setlength{\leftmargin}{#1}}%
         \item[]%
}
{\end{list}}

		\def\bsp{\begin{split}}
		\def\beq{\begin{eqnarray}}
		\def\bal{\begin{align*}}
		\def\bc{\begin{center}}
		\def\be{\begin{enumerate}}
		\def\bi{\begin{itemize}}
		\def\bs{\begin{small}}
		\def\bS{\begin{slide}}
		\def\ec{\end{center}}
		\def\ee{\end{enumerate}}
		\def\ei{\end{itemize}}
		\def\es{\end{small}}
		\def\eS{\end{slide}}
		\def\eeq{\end{eqnarray}}
		\def\eal{\end{align*}}
		\def\esp{\end{split}}
		\def\qed{ \vrule height7.5pt width7.5pt depth0pt}  

	\def\cp2problem#1#2#3#4{\fbox
		 {\begin{tabular*}{0.9\textwidth}
			{@{}l@{\extracolsep{\fill}}l@{\extracolsep{6pt}}l@{\extracolsep{\fill}}c@{}}
				#1 & & $#4 $ 
			\end{tabular*}}}

		\def\bkE{{\rm I\kern-.17em E}}
		\def\bk1{{\rm 1\kern-.17em l}}
		\def\bkD{{\rm I\kern-.17em D}}
		\def\bkR{{\rm I\kern-.17em R}}
		\def\bkP{{\rm I\kern-.17em P}}
		
		\def\bkZ{{\bf{Z}}}

\newcommand {\beeq}[1]{\begin{equation}\label{#1}}
\newcommand {\eeeq}{\end{equation}}
\newcommand {\bea}{\begin{eqnarray}}
\newcommand {\eea}{\end{eqnarray}}

\def\texitem#1{\par\smallskip\noindent\hangindent 25pt
               \hbox to 25pt {\hss #1 ~}\ignorespaces}

\def\bsp{\begin{split}}
		\def\beq{\begin{eqnarray}}
		\def\bal{\begin{align*}}
		\def\bc{\begin{center}}
		\def\be{\begin{enumerate}}
		\def\bi{\begin{itemize}}
		\def\bs{\begin{small}}
		\def\bS{\begin{slide}}
		\def\ec{\end{center}}
		\def\ee{\end{enumerate}}
		\def\ei{\end{itemize}}
		\def\es{\end{small}}
		\def\eS{\end{slide}}
		\def\eeq{\end{eqnarray}}
		\def\eal{\end{align*}}
		\def\esp{\end{split}}
		\def\qed{ \vrule height7.5pt width7.5pt depth0pt}  

\usepackage{amsmath, amssymb, xspace}
\usepackage{epsfig}
\usepackage{longtable}
\usepackage{color}
\usepackage{mathrsfs}
\usepackage{subfig}
\newenvironment{proof}[1][]{{\noindent \emph {Proof} #1: }}{\hfill \qed \vspace{3pt}\\ }

\title{Optimal Questionnaires for Screening of Strategic Agents}

\name{Anuj S. Vora, Ankur A. Kulkarni}
\address{Systems and Control Engineering, \\ Indian Institute of Technology Bombay, Mumbai, 400076}

\begin{document}
\maketitle

\section{Abstract}
\textit{[This is a longer version of our conference paper submitted to ICASSP 2021.]}
During the COVID-$19$ pandemic  the health authorities at airports and train stations try to screen and identify the travellers possibly exposed to the virus. However,  many individuals avoid getting tested  and hence may misreport their travel history. This is a challenge for the health authorities who wish to ascertain the truly susceptible cases in spite of this strategic misreporting. We investigate the problem of questioning travellers  to classify them for further testing when the travellers are strategic or are unwilling to reveal their travel histories. We show there are fundamental limits to how many travel histories the health authorities can recover.

\section{Introduction}


The  COVID-$19$ pandemic has demonstrated the need to screen and test travellers at airports and railway stations. However, due to limited resources, it is not feasible for the authorities to test all the travellers. Thus, it is imperative to identify the most susceptible travellers for testing by deriving information about their travel history and determining whether they have been to a COVID-$19$ hotspot. However, the travellers may not be willing to divulge their travel history to the relevant authorities. The question is then how to extract the true information from the travellers? In particular - given the tendency of the travellers to misreport their travel history, how should the health authorities query in order to screen and correctly identify maximum number of susceptible travellers?


The travellers have a predisposition to misrepresent their travel information. This could be  due to the associated stigma of the disease as well as the inconvenience caused due to the testing and quarantine protocols. These tendencies are different among each of the travellers. The health authorities thus face a wide spectrum of travellers and they have to \textit{strategize} appropriately so as to recover the maximum amount of information from the travellers.

We model this as a setting where a health inspector encounters travellers having an $n$-length travel history. A traveller can be classified baesd on its \textit{type} which determines its tendency to misreport its information. The health inspector wishes to know the travel history of the traveller but does not know the type of the traveller. Moreover, the traveller may have an incentive to misreport its travel history. We formulate this problem as a game between the traveller which we call the sender and the health inspector who is the receiver.

We show that the Stackelberg equilibrium strategy of the receiver, with the receiver as the leader, is to commit to a decoding function that meaningfully decodes only a subset of sequences correctly and deliberately induces an error on the rest of the sequences. In other words, the optimal strategy of the health inspector is to present a questionnaire with a predefined list of a subset of travel histories and ask the traveller to select any and exactly one.

We capture the misreporting nature of the sender with a utility that depends on the true sequence of travel history, the history recovered by the receiver and its type.  Crucial to our analysis is the notion of the \textit{sender graph} which is a graph on the space of the sequences induced by the utility of the sender. We define a notion of \textit{rate of information extraction} for the receiver which determines the growth of the perfectly recovered sequences with $n$. We compute upper and lower bounds on the maximum rate of information extraction. We show that barring the case when the types of the senders are completely divergent in their utility structures, the number of perfectly recovered sequences grows exponentially in $n$.

The problem of information extraction is related to the general problem of communication between sender and receiver with misaligned objectives studied in game theory \cite{crawford1982strategic, battaglini2002multiple, saritacs2015multi} control theory \cite{ sayin2019hierarchical, farokhi2016estimation, akyol2016information}, economics \cite{kamenica2011bayesian, bergemann2019information}. Our work is also related to the results in \cite{vora2020zero} where we discussed a case with a single sender and showed that the maximum rate of information extraction is bounded above by the Shannon capacity of a certain graph. In \cite{vora2020achievable, vora2020communicating}, we studied a related information extraction  problem where the receiver tried to achieve asymptotically vanishing probability of error. 

\section{Problem formulation}

\subsection{Notation}
The space of symbols is denoted by the calligraphic letter $\Xscr$ and the space of vector valued sequences is denoted as $\Xscr^n$. Note that to declutter notation, unless stated otherwise, the sequences $x, y, z$ will be vector valued.  The set of probability distributions on a space is denoted as $\Pscr(\cdot)$. For a graph $G$, the size of the largest independent set is denoted as $\alpha(G)$. The image of a function $g$ is denoted as $\image(g)$.

\subsection{Model}

Let $\Xscr$, $|\Xscr|< \infty$ be the set of all locations. A particular location $\Xscr$ may be a COVID-$19$ hotspot or a safe region. The travellers from these locations arrive at an airport or a train station with travel histories $x \in \Xscr^n$ which is a sequence of $n$ locations. Each of the travellers have a classification called the \textit{type} which determines their degree of honesty. We denote the type as $\lambda$ and the set of all types is denoted as $\Lambda$. We assume that $|\Lambda| < \infty$.

The health inspector wants to know the travel histories from each of the travellers. However, the inspector does not know the type of the traveller, but only has a noisy observation. We model this as a belief over the types denoted as $\Pbb_{\Lambda} \in \Pscr(\Lambda)$.

The inspector presents the travellers with a checklist which requires the travellers to choose anyone travel history from a specified  histories. Since the  inspector is unaware of the type of the traveller, the checklist is common across all travellers. If a  traveller with travel history $x$, states his history as $\xhat$, we say that the health inspector recovers a sequence $\xhat \in \Xscr^n$. The aim of the inspector is to get $\xhat = x$ without the knowledge of $x$ and $\lambda$. 


If the traveller is not honest, then it may wish to deceive the health inspector such that when $x$ is the travel history, then the inspector recovers $\xhat$ such that $\xhat \neq x$. We quantify this trait with a utility $\util_n(\xhat,x,\lambda)$ where $\util_n : \Xscr^n \times \Xscr^n \times \Lambda \rarr \Real$ is defined as
\begin{align}
    \util_n(\wi{x},x,\lambda) = \frac{1}{n} \sum_{i=1}^{n} \util(\wi{x}_i,x_i,\lambda) \qquad \forall \;x, \wi{x} \in \Xscr^n, \non
\end{align}
with $\util : \Xscr \times \Xscr \times \Lambda \rarr \Real$ being the single letter utility. Thus, the utility of a traveller depends on the history recovered by the health inspector $\xhat$, the true history $x$ and its type $\lambda$.

For a traveller with type $\lambda$ and history $x$, we say that the traveller \textit{prefers} $x'$ over $x$ if
 $ \util_n(x,x,\lambda) \leq \util_n(x',x,\lambda)$.
While $\util_n(x,x,\lambda) > \util_n(x'',x,\lambda)$ implies that the traveller does not prefer $x''$ over $x$. If $\util(x,x,\lambda) > \util(x',x,\lambda)$ for all $x,x' \in \Xscr, x' \neq  x$, then  we say that the traveller is honest. It also follows that $\util_n(x,x,\lambda) > \util_n(x',x,\lambda)$ for all $x, x' \in \Xscr^n, x' \neq x$ for the honest sender.

\subsection{Game formulation}

For the game formulation, we call the traveller as sender and the health inspector as a receiver. A sender with $\lambda \in \Lambda$ and travel history $x \in \Xscr^n $ responds to the checklist of the receiver as $s_n^\lambda(x) = y \in \Xscr^n$, where $s_n^\lambda : \Xscr^n \rarr \Xscr^n$. The receiver maps this response of the sender as $g_n(y) = \wi{x}$, where $g_n : \Xscr^n \rarr \Xscr^n$. Although this allows for a wide range of functions, we later show that without loss of generality, the receiver can choose a subset of sequences $I^n \subseteq \Xscr^n$ such that $g(x) = x$ for all $x \in I^n$ and $g(x') = \xbar$ for all $x' \notin I^n$, where $\xbar \in I^n$. In other words, it suffices that the health inspector need only specify a list of locations and ask the traveller to select exactly one.

Let $\Dscr(g_n,s_n^\lambda) :=  \curly{x \in \Xscr^n \;|\; g_n \circ s_n^\lambda(x) = x }$ be the set of perfectly recovered sequences when the receiver plays the strategy $g_n$ and the sender plays the strategy $s_n^\lambda$.  The receiver aims to maximize the size of the set $\Dscr(g_n,s_n^\lambda)$ averaged over the types of the senders by choosing an appropriate strategy $g_n$. The sender, on the other hand, tries to maximize the utility $\util_n(g_n \circ s_n^\lambda(x),x,\lambda)$ by choosing an appropriate strategy $s_n^\lambda$.  We formulate this problem as a game between the sender and receiver. In particular, we consider a leader-follower game, also called a \textit{Stackelberg} game, with the receiver as the leader and the sender as the follower. 

 The game proceeds as follows. The receiver, being the leader, plays and announces its strategy, \ie, a list of hitories, before the sender.  For a given strategy of the receiver, the sender chooses a response that maximizes its utility. The receiver anticipates this  response of the sender and  accordingly chooses an optimal strategy that maximizes its objective, \ie, a ``decoding'' function which, when composed with the sender's ``encoding'' strategy, recovers the maximum number of the sequences perfectly. This leads to the equilibrium concept called the \textit{Stackelberg} equilibrium solution \cite{basar99dynamic}. 
 \begin{definition}[Stackelberg equilibrium]\label{defn:stack-equi}
   In a Stackelberg equilibrium, the strategy of the receiver is given as
\begin{align}
g_n^* \in \arg \max_{g_n} \; \sum_{\lambda \in \Lambda} \Pbb(\lambda) \round{\min_{s_n^\lambda \in \best(g_n,\lambda)} |\Dscr(g_n,s_n^\lambda)|}, \label{eq:rec-opt-stra-game}
\end{align}
where the best response set of the sender, $\best(g_n,\lambda)$, is 
\begin{align}
  &\best(g_n,\lambda) = {\Big\{} s_n^\lambda : \Xscr^n \rarr \Xscr^n \;|\; \util_n(g_n \circ s_n^\lambda (x),x,\lambda) \geq \non \\
  & \hspace{1.5cm}\util_n(g_n \circ s_n'^\lambda (x),x,\lambda) \quad \forall \; x \in \Xscr^n, \forall \; s_n'^\lambda {\Big\}}. \label{eq:sen-opt-stra-game}
\end{align}   
 \end{definition}
This formulation is apt for our setting since the health inspector first presents the travellers with a checklist and then the travellers choose their responses accordingly.

 For a given strategy $g_n$ of the receiver, sender of each type responds with a strategy that depends on its utility structure. Thus, the set of best responses $\best(g_n,\lambda)$ is also a function of the type $\lambda$. In \eqref{eq:rec-opt-stra-game}, we minimize over the set of best responses of the sender $\best(g_n,\lambda)$ because the receiver does not have control over the choice of the best response of the sender. Thus, we assume that the receiver chooses its strategy according to the worst-case scenario and hence adopts a \textit{pessimistic} veiwpoint.

We now discuss an example which demonstrates the gameplay and some important aspects of game.
\begin{examp}\label{eg:two-types}
 Let $\Xscr = \{0,1,2\}$ and  $\Lambda = \{h,d\}$, where $h$ stands for \textit{honest} type and $d$ stands for \textit{dishonest} type. Let $\Pbb_\Lambda(h) = 1/3$ and $\Pbb_\Lambda(d) = 2/3$. Let the utility of the sender type $\lambda = h$ be 
$    \util(x,x,h) >     \util(x',x,h) \;\; \forall \; x' \in \Xscr, x' \neq x$. 
  Since $h$ is an honest sender, it does not prefer to lie about its history. The same is not true for the sender type $d$ which is a dishonest sender.  For the sender type $\lambda = d$, the utility  is
  \begin{align}
\begin{array}{c c c}
  \util(0,0,d) = 1,  &  \util(1,0,d) = 2, &  \util(2,0,d) = 0, \\
  \util(0,1,d) = 2, & \util(1,1,d) = 1, &  \util(2,1,d) = 0, \\
  \util(0,2,d) = 1, & \util(1,2,d) = 1, &  \util(2,2,d) = 0. 
\end{array}  \non
\end{align}


Thus, the dishonest sender prefers to lie about all its travel locations since $\util(x,x,d) < \util(x',x,d)$ for all $x' \neq x$.

Let $n = 1$. Suppose the receiver chooses a naive strategy $g : \Xscr \rarr \Xscr$ as $g(i) = i$ for all $i \in \Xscr$; in this strategy the receiver blindly believes the sender's word and maps it to the same location. Thus, its presents the sender with all possible options for reporting locations. The best response set of the sender type $h$ is the strategy $s^h(x) = x$ for all $x \in \Xscr$. For the sender type $d$, it can easily be checked that the best response set $\best(g,d) = \{\bar{s}^d,\wi{s}^d\}$, where
 \begin{align}
 \bar{s}^d(i) = \left\{
 \begin{array}{c l}
 1& i = 0 \\
 0& i = 1 \\
 0& i = 2
 \end{array}
 \right., \quad
 \wi{s}^d(i) = \left\{
 \begin{array}{c l}
 1& i = 0 \\
 0& i = 1 \\
 1& i = 2
 \end{array}
 \right.. \non
 \end{align}

  It can be observed the for the sender type $h$, the set of perfectly recovered sequences is $\Dscr(g,s^h) = \Xscr$. For the sender type $d$, $\Dscr(g,s^d) = \emptyset$ for all $s^d \in \best(g,d)$. This gives
  \begin{align}
\sum_{\lambda \in \Lambda} \Pbb(\lambda) \round{\min_{s^\lambda \in \best(g,\lambda)} |\Dscr(g,s^\lambda)|}  = \frac{1}{3}3 + \frac{2}{3}0 = 1.    \non
  \end{align}

  However, the receiver can do better by cleverly choosing its strategy.  Suppose instead the receiver chooses a strategy $\gtilde$ as $\gtilde(0) = 0 = \gtilde(1)$  and $\gtilde(2) = 2$.
    Thus, the receiver does not naively decode the sender's message. It chooses to correctly decode only a subset of sequences, in this case $\{0,2\}$ by applying an identity map on $\{0,2\}$. For the sequence $1$, the receiver decodes it to the sequence $0$ or equivalently, the checklist has only two locations $\{0,2\}$.

    Again, for the sender type $h$, the best response strategy is $\stilde^h(x) = x$ for all $x \in \Xscr$. For the sender type $d$, the best response strategy is $\stilde^d(i) = 0$ for all $i \in \Xscr$.  It can be observed that $\Dscr(g,\stilde^h) = \{0,2\}$ and $\Dscr(g,\stilde^d) = \{0\}$ and hence
          \begin{align}
            \sum_{\lambda \in \Lambda} \Pbb(\lambda) \round{\min_{s^\lambda \in \best(\gtilde,\lambda)} |\Dscr(\gtilde,s^\lambda)|}  = \frac{1}{3}2 + \frac{2}{3}1 = \frac{4}{3}.     \non
          \end{align}
                  Thus, a more sophisticated choice of strategy $\widetilde{g}$  improves on the naive strategy $g$. The difference between the strategies $g$ and $\widetilde{g}$ is that in the latter, the receiver \textit{deliberately} induces an error when it receives a sequence $1$ from the sender. Since $1$ is not in range of $\widetilde{g}$, sender has only $\{0,2\}$ as its choices. Given these choices, the sender type $d$ is forced to report $0$ as $0$. This is because, the sender does not prefer $2$ or over $0$ since $\util(2,0) < \util(0,0)$. Thus, the strategy $\widetilde{g}$ limits the choices of the sender and forces it to be honest for the sequence $0$. This strategy also limits the choices of the honest sender and thereby only $\{0,2\}$ are recovered. However, on average, the receiver does better since $\Pbb_\Lambda(h) < \Pbb_\Lambda(d)$.
                \end{examp}
                
\subsection{Some definitions}
In this section, we introduce some notions to quantify the amount of information that can be extracted from the senders. First, we define a notion of the sender graph.
\begin{definition} [\textit{Sender graph}] \label{defn:util-graph}
The sender graph for a sender of type $\lambda$, denoted as $G_\lambda^n = (\Xscr^n,E)$, is a graph where $(x,y) \in E$ if either
  $  \util_n(x,x,\lambda) \leq \util_n(y,x,\lambda)$  \newline or $ \util_n(y,y,\lambda) \leq \util_n(x,y,\lambda)$.
  
For $n = 1$, the graph $G_\lambda^1$ is denoted as $G_\lambda$.
\end{definition}
Thus, two sequences $x$ and $y$ are adjacent if the sender has an incentive to report one as the other.

\begin{definition}[Union of graphs]
  The union of two graphs $G_1 = (V_1,E_1)$ and  $G_2 = (V_2,E_2)$ is defined as a graph $G = (V,E)$ where $V = V_1 \cup V_2$ and $E = E_1 \cup E_2$.
\end{definition}

\begin{definition}[$\lambda$-partition of a set] \label{defn:ihat-ibar}
Let $I^n \subseteq \Xscr^n$ be any set. For $\lambda \in \Lambda$, the $\lambda$-partition of the set $I^n$ is defined as 
    \begin{equation}
 \begin{aligned}
   \Ibar_\lambda^n &:= \{x \in I^n : \util_n(x,x,\lambda) > \util_n(y,x,\lambda) \;\forall \; y \in I^n, y \neq x\}. \non
 \end{aligned}
\end{equation}
For the set of types $\Lambda$, let the collection of all $\lambda$-partitions of the set $I^n$ be denoted as
$  \Fscr(I^n) = \{\Ibar_\lambda^n\}_{ \lambda \in \Lambda}$.

\end{definition}
Thus, the set $\Ibar_\lambda^n$  is such that when the sender observes a sequence $ x \in \Ibar_\lambda^n$, the sender prefers to report $x$ honestly amongst the sequences in $I^n$.


We now define the notion of rate which determines the growth of the perfectly recovered sequences with $n$.
\begin{definition}[Rate of information extraction] \label{defn:rate}
  For a strategy $g_n$ of the receiver, define $\Dscr^*(g_n)$
  \begin{align}
\Dscr^*(g_n) =  \sum_{\lambda \in \Lambda} \Pbb(\lambda) \round{\min_{s_n^\lambda \in \best(g_n,\lambda)} |\Dscr(g_n,s_n^\lambda)|}. \non
  \end{align}
  Then, the rate of information extraction is defined as
  \begin{align}
  R(g_n) = \round{ \Dscr^*(g_n)}^{1/n}. \non
 \end{align}
  
\end{definition}

\section{Results}


We now characterize the set of perfectly recovered sequences in a Stackelberg equilibrium of the game.
  \begin{theorem} \label{thm:Ibar_Ihat_recovery}
    Let $n \in \Nbb$ be fixed. For a sender type $\lambda \in \Lambda$ with utility $\util_n(\cdot,\cdot,\lambda)$, let $G_\lambda^n$ be the corresponding sender graph. Let $g_n$ be any strategy of the receiver and let $\Fscr(\image(g_n)) = \{\Ibar_\lambda^n\}_{\lambda \in \Lambda}$ be the collection of the $\lambda$-partitions  of the set $\image(g_n)$.  Then, 
     \begin{align}
\Dscr^*(g_n)  = \sum_{\lambda \in \Lambda} \Pbb(\lambda) |\Ibar_\lambda^n|.  \non
  \end{align}
 \end{theorem}
 \begin{proof}
 Consider the sender with type $\lambda$. Let $\xbar \in \Ibar_\lambda^n$. Then, from the definition of $\Ibar_\lambda^n$, we have 
$          \util_n(\xbar,\xbar,\lambda) > \util_n(x',\xbar,\lambda) \quad \forall \; x' \in I^n, x' \neq  \xbar$. 
     Thus, for all $s_n^\lambda \in \best(g_n,\lambda)$, $g_n \circ s_n^\lambda (\xbar) = \xbar$. Since for all other $s_n'^\lambda $ where $g_n \circ s_n'^\lambda(\xbar) = x'$, we have
     \begin{align}
       \util_n(g_n \circ s_n'^\lambda(\xbar),\xbar,\lambda)  &= \util_n(x',\xbar,\lambda) \non \\
       &\leq \util_n(\xbar,\xbar,\lambda) = \util_n(g_n \circ s_n^\lambda(\xbar),\xbar,\lambda), \non
     \end{align}
and hence
  $     \util_n(g_n \circ s_n'^\lambda(\xbar),\xbar,\lambda)  \leq \util_n(g_n \circ s_n^\lambda(\xbar),\xbar,\lambda)       \;\; \forall \; s_n'^\lambda$,
     with equality if and only if $g_n \circ s_n'^\lambda(\xbar) = g_n \circ s_n^\lambda(\xbar)$. Thus, for all $s_n^\lambda \in \best(g_n,\lambda)$, we have $g_n \circ s_n^\lambda(\xbar) = \xbar$.
     
Let $\Ihat_\lambda^n = I^n \setminus \Ibar_\lambda^n$. Then, from the definitions of $\Ibar_\lambda^n$, we have that for all $\yhat \in \Ihat_\lambda^n$,  there exists an $y' \in I^n, y' \neq \yhat$ such that $ \util_n(\yhat,\yhat,\lambda) \leq  \util_n(y',\yhat,\lambda)$.     Thus, there exists $s_n'^\lambda \in \best(g_n,\lambda)$ such that $g_n \circ s_n'^\lambda(\yhat) \neq \yhat$ for all $\yhat \in \Ihat_\lambda^n$.

Thus, all $\xbar \in \Ibar_\lambda^n$ are recovered by the receiver irrespective of the strategy chosen by the sender. However, in the worst case no sequence in $\Ihat_\lambda^n$ is recovered. Thus, 
   $     \min_{s_n^\lambda \in \best(g_n,\lambda)} |\Dscr(g_n,s_n^\lambda)| = |\Ibar_\lambda^n|$. 
This result holds for all senders and hence we get
 $\Dscr^*(g_n) = \sum_{\lambda \in \Lambda} \Pbb(\lambda) |\Ibar_\lambda^n|$.
   \end{proof}
   This theorem show that for a strategy $g_n$ of the receiver, the perfectly recovered sequences are given by the set $\cup_\lambda \Ibar_\lambda^n$, where $\Ibar_\lambda^n$ is the $\lambda$-partition of $\image(g_n)$. However, given a set $I^n$ and its collection $\lambda$-partition $\Fscr(I^n)$, it is not clear how to perfectly recover the set $\cup_\lambda \Ibar_\lambda^n$. The proof of the following theorem suggests one such strategy for the receiver. We also characterize the set of perfectly recovered sequences in a Stackelberg equilibrium of the game. 

   \begin{theorem}
    Let $n \in \Nbb$ be fixed. For a sender type $\lambda \in \Lambda$ with utility $\util_n(\cdot,\cdot,\lambda)$, let $G_\lambda^n$ be the corresponding sender graph.
   Then,
         for all equilibrium strategies $g_n^*$ of the receiver 
     \begin{align}
\Dscr^*(g_n^*)  = \max_{I^n \subseteq \Xscr^n}\sum_{\lambda \in \Lambda} \Pbb(\lambda) |\Ibar_\lambda^n|, \non
     \end{align}
     where $\{\Ibar_\lambda^n\}_{\lambda \in \Lambda}$ is the collection of the $\lambda$-partitions of $I^n$.
   \end{theorem}
   \begin{proof}
     Let $I^n \subseteq  \Xscr^n$ be any set and let  $\Ibar_\lambda^n$ be its $\lambda$-partition. Define a strategy $g_n$ for the receiver as
        \begin{align}
     g_n(x) = \left\{
     \begin{array}{c l}
       x & x \in I^n \\
       \xbar & x \notin I^n
     \end{array}
\right., \label{eq:g_n-defn}
        \end{align}
        where $\xbar$ is some sequence in $\Ibar_\lambda^n$.                             From this choice of $g_n$ is it clear that $\image(g_n) = I^n$. From Theorem~\ref{thm:Ibar_Ihat_recovery}, we get that
$\Dscr^*(g_n)  = \sum_{\lambda \in \Lambda} \Pbb(\lambda) |\Ibar_\lambda^n|$. 
                            Maximizing over the choice of strategies $g_n$ is equivalent to maximizing over $I^n$ and hence
                            \begin{align}
\max_{g_n} \Dscr^*(g_n) =  \Dscr^*(g_n^*)  = \max_{ I^n \subseteq \Xscr^n}\sum_{\lambda \in \Lambda} \Pbb(\lambda) |\Ibar_\lambda^n|. \non
 \end{align}
                             \vspace{-0.3cm}                          
\end{proof}
Thus, we see that the optimal questionnaire consists only a subset of histories $I^n$ from $\Xscr^n$. 
Thus, to determine the optimal questionnaire, the receiver has to determine the largest set $I^n$ for which $\sum_{\lambda \in \Lambda} \Pbb(\lambda) |\Ibar_\lambda^n|$ is maximized. A particular choice of the set $I^n$ would be to take the largest independent set in the graph $\cup_\lambda G_\lambda^n$, denoted as $\Itilde^n$. Since $\Itilde^n$ is also an independent set in each of the sender graphs $G_\lambda^n$, we get that $\Ibar_\lambda^n = \Itilde^n$ for all $\lambda \in \Lambda$. Thus, $\Dscr(g_n) = \alpha(\cup_\lambda G_\lambda^n)$. However, in the following we show that set of set may not be the optimal choice for the receiver.

   \begin{corollary}\label{coro:bounds-on-rate}
Let $n \in \Nbb$ be fixed. For a sender type $\lambda \in \Lambda$ with utility $\util_n(\cdot,\cdot,\lambda)$, let $G_\lambda^n$ be the corresponding sender graph. Then, for all Stackelberg equilibrium strategies $g_n^*$,
     \begin{align}
\alpha\round{\cup_{\lambda}G_\lambda^n}^{1/n} \leq R(g_n^*) \leq \round{\sum_{\lambda \in \Lambda} \Pbb_\Lambda(\lambda) \alpha(G_\lambda^n)}^{1/n}. 
  \end{align}
\end{corollary}
\begin{proof}
  Let $g_n$ be a strategy of the receiver such that $|\image(g_n)| = \alpha\round{\cup_{\lambda}G_\lambda^n}$. Define $I^n = \image(g_n)$. Clearly, $I^n$ is an independent set in $G_\lambda^n$ and hence for all $\lambda$, the $\lambda$-partition is $\Ibar_\lambda^n = I^n$. This gives that
 $   \Dscr^*(g_n) = \alpha\round{\cup_{\lambda}G_\lambda^n} \leq     \Dscr^*(g_n^*)$.
  Thus,
    $R(g_n^*) \geq     \Dscr^*(g_n)^{1/n} = \alpha\round{\cup_{\lambda}G_\lambda^n}^{1/n.}$.
  For the upper bound, we use the fact that $\Ibar_\lambda^n$ is an independent set in $G_\lambda^n$. Thus, $|\Ibar_\lambda^n| \leq \alpha(G_\lambda^n)$, which gives that
$ R(g_n^*) =  \Dscr^*(g_n^*)^{1/n} \leq \round{\sum_{\lambda \in \Lambda} \Pbb_\Lambda(\lambda)\alpha(G_\lambda^n)}^{1/n}$. \non
\end{proof}
\begin{remark}
In the Example~\ref{eg:two-types}, only singleton sets are independent sets in $G_h \cup G_d$ and hence $\alpha(G_h \cup G_d) = 1$. Thus, if the receiver chooses a strategy $g(x) = 0 \;\forall \; x \in \Xscr$, then $\Dscr(g,s^\lambda) = \{0\} \; \forall \;s^\lambda \in \best(g,\lambda), \; \forall \; \lambda \in \Lambda$. This gives $R(g)  = 1$. However, as demonstrated in Example~\ref{eg:two-types}, this choice of strategy is sub-optimal for the receiver.
\end{remark}

We now take the limit as $n \rarr \infty$ and derive fundamental bounds on the rate of information extraction. 


\begin{theorem} \label{thm:lim-sup-bound}
Consider senders having type $\lambda \in \Lambda$ with utility $\util(\cdot,\cdot,\lambda)$ and  let $\{G_\lambda^n\}_{n \geq 1}$ be the sequence of sender graphs. Then, for all sequences of Stackelberg equilibrium strategies $\{g_n^*\}_{n \geq 1}$  of the receiver,
     \begin{align}
 \alpha\round{\cup_{\lambda}G_\lambda} \leq \lim \sup_{n \rarr \infty} R(g_n^*) \leq  \Xi(\util,\lambda^*).  \label{eq:max-rate-bounds}
     \end{align}
     where $\lambda^* = \max_{\lambda \in \Lambda} \alpha(G_\lambda)$ and \newline$\Xi(\util,\lambda^*) := \lim_{n \rarr \infty} \alpha(G_{\lambda^*}^n)^{1/n}$.
   \end{theorem}
   \begin{proof}
We first derive the lower bound.       Let $I$ be the largest independent set in $\cup_\lambda G_\lambda$, \ie, $|I^n| = \alpha\round{\cup_\lambda G_\lambda}$. Let $I^n := I \times \hdots \times I$. We show that $I^n$ is an independent set in the graph $\cup_\lambda G_\lambda^n$. Let $x,y \in I^n$ be distinct sequences and consider the difference of utilites for type $\lambda$
  \begin{align}
    &\util_n(x,x,\lambda) - \util_n(y,x,\lambda) \non \\
    &= \frac{1}{n}\sum_{i=1}^{n}    \util(x_i,x_i,\lambda) - \util(y_i,x_i,\lambda). \non
  \end{align}
  Since $x_i, y_i \in I$ for all $i \in \{1,\hdots,n\}$, we have \newline$\util(x_i,x_i,\lambda) > \util(y_i,x_i,\lambda)$ for all $i$. Thus, $ \util_n(x,x,\lambda) > \util_n(y,x,\lambda)$. We can similarly prove that $  \util_n(y,y,\lambda) > \util_n(x,y,\lambda)$. This holds for all $\lambda \in \Lambda$ and hence  $x, y \in I^n$ are non-adjacent in the graph $\cup_\lambda G_\lambda^n$. Thus,  
  \begin{align}
\alpha \round{\cup_\lambda G_\lambda^n} \geq |I^n| = \alpha(\cup_\lambda G_\lambda)^n. \non
  \end{align}
  Using the lower bound in Corollary~\ref{coro:bounds-on-rate} and taking the limit superior as $n \rarr \infty$, the lower bound in \eqref{eq:max-rate-bounds} follows.

      We now derive the upper bound. We have \newline $\round{\sum_{\lambda \in \Lambda} \Pbb_\Lambda(\lambda) \alpha(G_\lambda^n)}^{1/n}$
     \begin{align}
       &= \round{\Pbb_\Lambda(\lambda^*) \alpha(G_{\lambda^*}^n)}^{1/n}  \round{ 1 + \frac{\sum_{\lambda \in \Lambda, \lambda \neq \lambda^*} \Pbb_\Lambda(\lambda) \alpha(G_\lambda^n)}{\Pbb_\Lambda(\lambda^*) \alpha(G_{\lambda^*}^n)}}^{1/n}. \non
     \end{align}
Using $\alpha(G_{\lambda^*}) \geq \alpha(G_\lambda)$, it can be shown that for all $n$,  $\alpha(G_{\lambda^*}^n) \geq \alpha(G_\lambda^n)$ \cite{vora2020information}. Thus,
     \begin{align}
\lim_{n \rarr \infty}\frac{\sum_{\lambda \in \Lambda, \lambda \neq \lambda^*} \Pbb_\Lambda(\lambda) \alpha(G_\lambda^n)}{\Pbb_\Lambda(\lambda^*) \alpha(G_{\lambda^*}^n)} = 0 \non
     \end{align}
     and hence 
     \begin{align}
\lim_{n \rarr \infty}        \round{ 1 + \frac{\sum_{\lambda \in \Lambda, \lambda \neq \lambda^*} \Pbb_\Lambda(\lambda) \alpha(G_\lambda^n)}{\Pbb_\Lambda(\lambda^*) \alpha(G_{\lambda^*}^n)}}^{1/n} = 1. \non
     \end{align}
     Using this and $\Pbb_\Lambda(\lambda^*)^{1/n} \rarr 1$, we get
     \begin{align}
\lim_{n \rarr \infty}        &\round{\sum_{\lambda \in \Lambda} \Pbb_\Lambda(\lambda) \alpha(G_\lambda^n)}^{1/n}   = \lim_{n \rarr \infty} \alpha(G_{\lambda^*}^n)^{1/n}. \label{eq:lim-rhs}
     \end{align}
     The limit on the right hand side exists \cite{vora2020information} and is denoted as $\Xi(\util,\lambda^*)$. Using the upper bound in Corollary~\ref{coro:bounds-on-rate} and taking the limit superior as $n \rarr \infty$, the upper bound in \eqref{eq:max-rate-bounds} follows.
   \end{proof}
\bibliographystyle{IEEEbib}
\bibliography{ref.bib}

\end{document}